\documentclass[aps,prl,twocolumn,amsmath,longbibliography]{revtex4}  
\usepackage{bbm}
\usepackage{mathrsfs}
\usepackage{amsmath}
\usepackage{amsfonts}
\usepackage[colorlinks=true,citecolor=blue,anchorcolor=blue]{hyperref}
\usepackage{graphicx,epstopdf}
\usepackage{subfigure}
\usepackage{epsfig}
\usepackage{dcolumn}
\usepackage{bm}
\usepackage{color}
\usepackage{natbib}
\usepackage{amssymb}
\usepackage{xcolor}
\usepackage{braket}

\begin{document}

\title{Tunable dynamical magnetoelectric effect in antiferromagnetic topological insulator MnBi$_2$Te$_4$ films}

\author{Tongshuai Zhu$^{1,*}$, Huaiqiang Wang$^{1,*}$, Haijun Zhang$^{1,2,\dagger}$ and Dingyu Xing$^{1,2}$}
\affiliation{
$^1$ National Laboratory of Solid State Microstructures, School of Physics, Nanjing University, Nanjing 210093, China\\
$^2$ Collaborative Innovation Center of Advanced Microstructures, Nanjing University, Nanjing 210093, China\\
}

\begin{abstract}
More than forty years ago, axion was postulated as an elementary particle with a low mass and weak interaction in particle physics to solve the strong $\mathcal{CP}$ (charge conjugation and parity) puzzle. Axions are also considered as a possible component of dark matter of the universe. However, the existence of axions in nature has not been confirmed. Interestingly, axions arise as pseudoscalar fields derived from the Chern-Simons theory in condensed matter physics. In antiferromagnetic insulators, the axion field can become dynamical induced by spin-wave excitations and exhibits rich exotic phenomena, such as, the chiral magnetic effect, axionic polariton and so on. However, the study of the dynamical axion field is rare due to the lack of real materials. Recently, MnBi$_2$Te$_4$ was discovered to be an antiferromagnetic topological insulator with a quantized axion field protected by the inversion symmetry $\mathcal{P}$ and the magnetic-crystalline symmetry $\mathcal{S}$. Here, we studied MnBi$_2$Te$_4$ films in which both the $\mathcal{P}$ and $\mathcal{S}$ symmetries are spontaneously broken and found that the dynamical axion field and largely tunable dynamical magnetoelectric effects can be realized through tuning the thickness of MnBi$_2$Te$_4$ films, the temperature and the element substitution. Our results open a broad avenue to study axion dynamics in antiferromagnetic topological insulator MnBi$_2$Te$_4$ and related materials, and also is hopeful to promote the research of dark matter.

\end{abstract}

\email{zhanghj@nju.edu.cn}

\maketitle

\begin{figure}[t]
  \centering
  \includegraphics[width=3.4in]{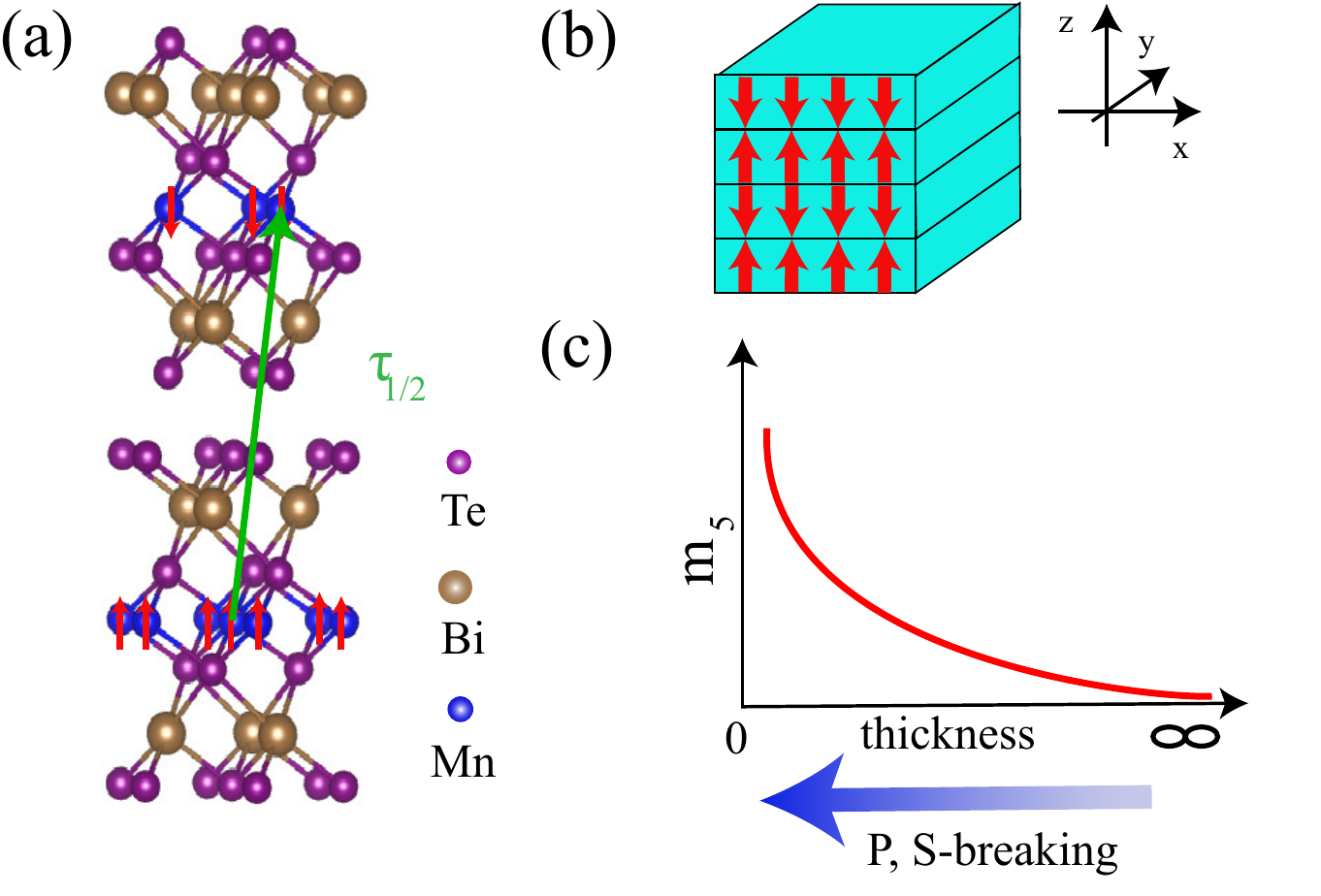}\\
  \caption{Schematic of $\mathcal{P}$, $\mathcal{S}$-breaking mass in  MnBi$_2$Te$_4$ thin films. (a) Crystal structure of MnBi$_2$Te$_4$. The red arrows denote the spin moment of Mn atoms and the green arrows represent the half-translation operator $\tau_{1/2}$ for the $\mathcal{S}$ symmetry. (b) Schematic of MnBi$_2$Te$_4$ thin films with even septuple layers grown along the easy axis (the $z$-direction). Both the inversion symmetry $\mathcal{P}$ and $\mathcal{S}$ symmetry are explicitly broken. (c) Schematic of the $\mathcal{P}$, $\mathcal{S}$-breaking mass $m_5$ in MnBi$_2$Te$_4$ thin films. It expects to increase when the thickness of thin films decreases due to the finite-size effect. }\label{fig1}
\end{figure}

In condensed matter physics, an effective axion field can be derived from the (4+1)-dimensional Chern-Simons theory of topological insulators(TIs) \cite{qi2011_rmp,qi2008_prb}, described by an axion action $S_{\theta}=\frac{\theta}{2\pi}\frac{e^2}{h}\bm{E}\cdot{\bm{B}}$, in which $\bm{E}$ and $\bm{B}$ are the electromagnetic fields inside the insulators, $e$ is the charge of an electron, $h$ is Plank's constant, $\theta$ is the dimensionless pseudoscalar parameter as the axion field \cite{wilczek1987}. The axion field $\theta$ is odd under the time reversal symmetry $\mathcal{T}$  or spatial inversion symmetry $\mathcal{P}$, so it is quantized to $\pi$ (mod $ 2\pi$) for TIs and 0 (mod $ 2\pi$) for normal insulators (NIs) if $\mathcal{T}$ or $\mathcal{P}$ is preserved. Such a quantized $\theta$ can lead to the image magnetic monopole effect \cite{qi2009monopole}, the quantized magneto-optical Faraday/Kerr effect \cite{maciejko2010,tse2010,tetsuyuki2012,karch2009,malshukov2013,wu2016,okada2016,dziom2017}, the topological magnetoelectric effect (TME) \cite{qi2008_prb,essin2009,rosenberg2010,coh2011,morimoto2015,wang2015,zirnstein2017} and the half-integer quantum Hall effect on the $\mathcal{T}$-breaking surface of TIs \cite{qi2008_prb}. Recently, the axion field in condensed matters was reported in charge-density-wave Weyl semimetals\cite{gooth2019}, magnetic topological insulators\cite{mong2010,wan2012,zhang2019mbt,liu2020robust,xu2019axion,turner2012} and topological-material heterostructures\cite{mogi2017,xiao2018}. More detailed discussions of the axion physics in condensed matters can be found in a recent review\cite{nenno2020}.

\begin{figure*}[t]
  \centering
  \includegraphics[width=5.0in]{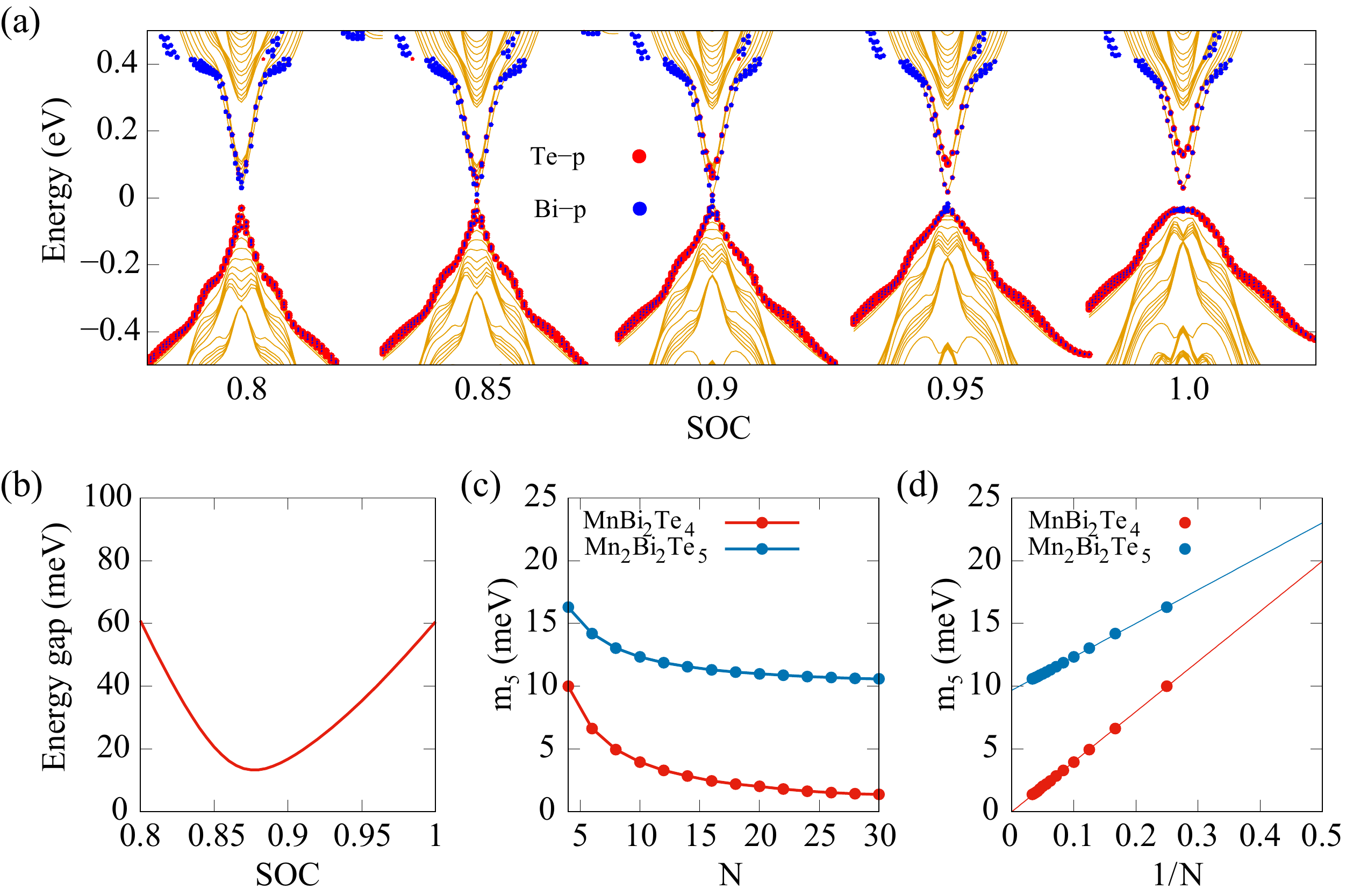}\\
  \caption{ Finite-size effect of MnBi$_2$Te$_4$ thin films. (a) The evolution of the band structure of the 12-SL MnBi$_2$Te$_4$ thin film as a function of the strength of the spin-orbit coupling (SOC). The energy gap first goes smaller and then goes larger without exactly closing. (b) The evolution of the energy gap as the SOC strength. It is clear to see that there is a minimum energy gap when increasing the SOC. The minimum energy gap corresponds to $2|m_5|$ (the $\mathcal{P}$, $\mathcal{S}$-breaking mass). (c,d) The mass $m_5$ versus the thickness of MnBi$_2$Te$_4$ (red) and Mn$_2$Bi$_2$Te$_5$ (blue) thin films. $N$ presents the number of bi-SLs of MnBi$_2$Te$_4$ films or (nonuple layers)NLs of Mn$_2$Bi$_2$Te$_5$ films. }\label{fig2}
\end{figure*}

In antiferromagnetic (AFM) insulators, the axion field $\theta$ can be a non-quantized value, once both $\mathcal{T}$ and $\mathcal{P}$ are broken. The spin-wave excitations can induce fluctuations of the axion field\cite{li2010dynamical,wang2011prl,zhang2020cpl}, called as the dynamical axion field $\theta(\bm{r},t)$ (DAF) which has spatial and temporal dependence. $\theta(\bm{r},t)$ can lead to rich dynamical magnetoelectric (ME) effects, for example, the dynamical chiral magnetic effect\cite{sekine2016,sumiyoshi2016,li2016,sekine2016prb}, anomalous Hall effect \cite{sekine2016}, axionic polariton \cite{li2010dynamical} and nonlinear electromagnetic effect \cite{ooguri2012,imaeda2019jpsj}. In principle, generic AFM insulators may exhibit the DAFs, such as Cr$_2$O$_3$ \cite{hehl2008,coh2011}, but the DAFs turn out to be too weak to be detected \cite{zhang2020cpl}. Large DAF requires the topologically nontrivial AFM insulators hosting nonzero spin Chern numbers \cite{wang2020dynamical}. Recently, van der Waals(vdW) layered material Mn$_2$Bi$_2$Te$_5$ and the superlattice (MnBi$_2$Te$_4$)$_2$(Bi$_2$Te$_3$)$_1$ were predicted to host large DAFs \cite{zhang2020cpl,wang2020dynamical}, denoted as DAF insulators. However, though material Mn$_2$Bi$_2$Te$_5$ has been successfully grown in experiments, the sample quality needs to be further improved. The experimental synthesis of the superlattice (MnBi$_2$Te$_4$)$_2$(Bi$_2$Te$_3$)$_1$ has not been reported. Therefore, the urgent challenge is to discover DAF insulators.

Recently, bulk MnBi$_2$Te$_4$ was discovered to be an AFM topological insulator and it hosts a quantized axion field $\theta=\pi$ protected by the inversion symmetry $\mathcal{P}$ and the magnetic-crystalline symmetry $\mathcal{S}=\mathcal{T}\tau_{1/2}$, where $\tau_{1/2}$ is the half translation operator \cite{gong2019cpl,zhang2019mbt,li2019sa,otrokov2019nature,deng2020,liu2020robust,chen2019intrinsic}. Interestingly, both $\mathcal{P}$ and $\mathcal{S}$ are spontaneously broken in MnBi$_2$Te$_4$ films with even septuple (SL) layers. In this work, our results indicate that the fluctuations of AFM order can lead to a dynamical ME effect within a large tunable range in MnBi$_2$Te$_4$ thin films, induced by the $\mathcal{P}$ and $\mathcal{S}$-breaking mass $m_5$.

As shown in Fig.~\ref{fig1}a, the antiferromagnetic topological insulator MnBi$_2$Te$_4$ \cite{zhang2019mbt} consists of  SLs coupled to each other by a Van der Waals-type interaction. It features an A-type AFM magnetic ground state with the out-of-plane easy axis. Although the magnetic order explicitly breaks the time-reversal symmetry $\mathcal{T}$,  the bulk MnBi$_2$Te$_4$ possesses a combined magnetic-crystalline symmetry $S=\mathcal{T}\tau_{1/2}$, marked in Fig.~\ref{fig1}a. In addition,  the inversion symmetry $\mathcal{P}$ is preserved for the bulk MnBi$_2$Te$_4$ with the inversion center at the Mn layer in the middle of each SL. Interestingly,  both the $S$ symmetry and the inversion symmetry $\mathcal{P}$ are explicitly broken in even-SL MnBi$_2$Te$_4$ thin films due to the finite size, schematically shown in Fig.~\ref{fig1}b. As a result, the restriction for the quantization of the axion filed $\theta$ is removed, which will lead to a tunable dynamical ME effect as we will demonstrate in details.

\emph{Effective model analysis.} The topological nature and the physical properties of MnBi$_2$Te$_4$ are determined by the electronic structure near the $\Gamma$ point because the minimum energy gap is at the $\Gamma$ point.  Therefore, we consider the effective $k\cdot p$ model with the four low-lying states $\ket{P1^+_z,\uparrow(\downarrow)}$ and $\ket{P2^-_z,\uparrow(\downarrow)}$ at the $\Gamma$ point, where `$\uparrow(\downarrow)$' indicates the spin. The state $\ket{P1^+_z}$ stands for the bonding states of two Bi layers and $\ket{P1^+_z}$ for the anti-bonding state of two Te layers. We start from the low-energy effective Hamiltonian of bulk MnBi$_2$Te$_4$ which hosts the symmetries $\mathcal{S}$, $\mathcal{P}$ and $\mathcal{C}_{3z}$. The four-band $k\cdot p$ model up to the third order of $k$ is written down as ~\cite{zhang2019mbt},

\begin{equation}
\label{dkgamma}
H=H_{0}+H_{3},
\end{equation}
where 
\begin{equation}
H_0=\epsilon(\bm{k})\Gamma_0+\sum_{i=1}^5d_i(\bm{k})\Gamma^i,
\end{equation}
with $d_i(\bm{k})=\Big(A_2k_y,-A_2k_x,A_1k_z,M(\bm{k}),0\Big)
$, $\epsilon(\bm{k})=C_0+D_1k_z^2+D_2(k_x^2+k_y^2)$, and $M(\bm{k})=m_4+B_1k_z^2+B_2(k_x^2+k_y^2)$, where $m_4$ is the mass term $m_4<0$ ($m_4>0$) for with (without) the band inversion in the band structure, and the Dirac matrices are $\Gamma_{0,1,...,5}=(\tau_0\otimes\sigma_0, \tau_x\otimes\sigma_x,\tau_x\otimes\sigma_y, \tau_x\otimes\sigma_0, \tau_z\otimes\sigma_0, \tau_x\otimes\sigma_z)$, where $\tau_{x,y,z}$ and $\sigma_{x,y,z}$ denote the Pauli matrices for the orbital and spin, respectively. The third order term $H_3$ is given explicitly by
\begin{equation}
H_3=R_{1}\left(k_{x}^{3}-3 k_{x} k_{y}^{2}\right) \Gamma_{5}+R_{2}\left(3 k_{x}^{2} k_{y}-k_{y}^{3}\right) \Gamma_{3}.
\end{equation}
The detailed parameters can be obtained by fitting with first-principles band structure and have been provided in Ref. \cite{zhang2019mbt}, as will be adopted here.

\begin{figure}[t]
  \centering
  \includegraphics[width=3.4in]{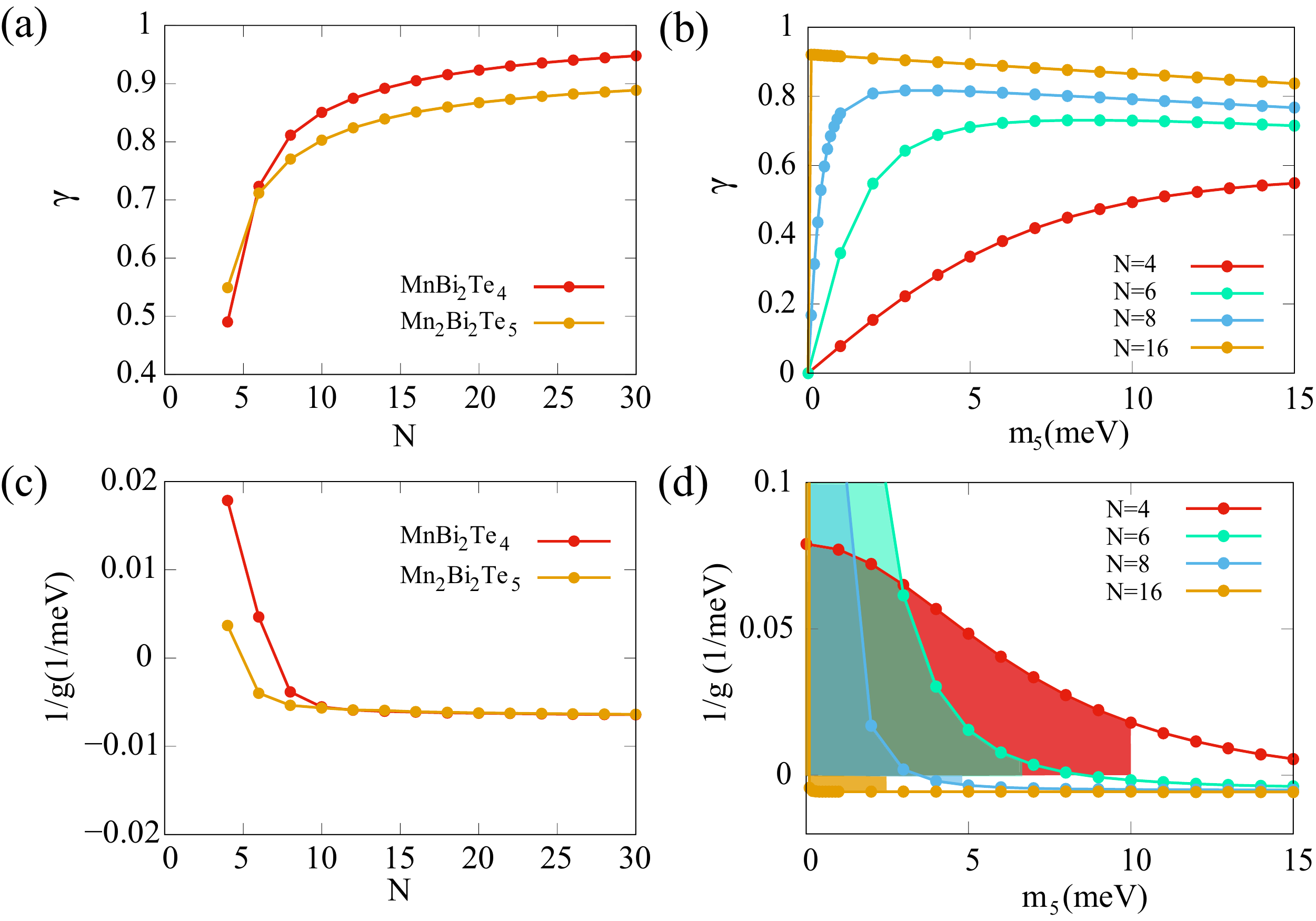}\\
  \caption{Dynamical magnetoelectic effect. (a) The magnetoelectric quantity $\gamma$ and (b) $1/g$ versus the thickness with $N$ bi-SLs for MnBi$_2$Te$_4$ (red) and NLs Mn$_2$Bi$_2$Te$_5$ (yellow) thin films, where realistic values of $m_5$ are obtained from first-principles calculations. (c,d) $\gamma$ (c) and $1/g$ (d) as a function of $m_{5}$ for films with different thickness ($N=4, 6, 8, 16$).}\label{fig3}
\end{figure}

Further for MnBi$_2$Te$_4$ thin films, both the $\mathcal{S}$ symmetry and inversion symmetry $\mathcal{P}$ are broken, but the combined symmetry $\mathcal{PT}$ is preserved. Based on the symmetry analysis, a mass term $m_5\Gamma_5$ to the leading order adds to the effective $k\cdot p$ model (Eq.~\ref{dkgamma}) for MnBi$_2$Te$_4$ thin films. The mass term $m_5\Gamma_5$ can significantly change the axion field $\theta$ from the quantized value ($\pi$). Actually, the mass term $m_5$ results from the interplay between the AFM order and the finite-size effect, and it is expected to decreases with increasing thickness of the films, and finally vanishes for the bulk limit, as illustrated in Fig.~\ref{fig1}c. This is in contrast to the three-dimensional (3D) DAF insulator~\cite{li2010dynamical, zhang2020cpl, wang2020dynamical} with a non-vanishing $m_5$ term in the bulk limit. Based on the effective $k\cdot p$ model (Eq.~\ref{dkgamma}), the energy gap at the $\Gamma$ point of MnBi$_2$Te$_4$ thin films is found to be $E_{gap}=2\sqrt{m_4^2+m_5^2}$. When the strength of spin-orbit coupling (SOC) is tuned from zero to the realistic value, the band inversion is induced along with the mass term $m_4$ changing from $m_4>0$ to $m_4<0$, and the energy gap $E_{gap}=2|m_5|$ reaches the minimum when the band inversion starts ($m_4=0$). Based on this consideration, we can quantitatively characterize the mass $m_{5}$ of MnBi$_2$Te$_4$ thin films.

As an example, the fat band structures of 12-SL MnBi$_2$Te$_4$ with different SOC are calculated and shown in Fig.~\ref{fig2}a. When SOC is weak, the band structure indicates an energy gap $E_{gap}$ without the band inversion. As expect, when increasing SOC, $E_{gap}$ becomes smaller, and then reaches the minimum around SOC$\sim 0.9$, and then becomes bigger with the band inversion. The SOC dependence of $E_{gap}$ is further extracted and shown in Fig.~\ref{fig2}b.  In Fig.~\ref{fig2}c, it explicitly shows the mass $m_5$ as a function of the thickness of $N$ bi-SL MnBi$_2$Te$_4$ thin films (red line), which indeed gradually decreases with increasing $N$. For the comparison, we also present the result of $m_5$ for Mn$_2$Bi$_2$Te$_5$ thin films (blue line). It is obvious that for large $N$ towards the bulk limit, $m_5$ approaches zero (a finite value) for MnBi$_2$Te$_4$ (Mn$_2$Bi$_2$Te$_5$), as expected. It is worth to notice that $m_5$ of MnBi$_2$Te$_4$ films is tunable to zero within a larger range than that of Mn$_2$Bi$_2$Te$_5$ films, as shown in Fig.~\ref{fig2}c. Furthermore, it is numerically found that $m_5$ exhibits an approximately linear behavior against $1/N$, namely, the inverse of the film thickness for both MnBi$_2$Te$_4$ and Mn$_2$Bi$_2$Te$_5$ films, as shown in Fig.~\ref{fig2}d. Our results provide a practical guideline to tune the mass term $m_5$ through varying the thickness of MnBi$_2$Te$_4$ films.

\emph{Dynamical magnetoelectric effect.} The application of an electric field can induce a magnetization, and reversely, the application of a magnetic field can lead to a polarization, which is the typical ME effect. In 3D topological insulators, such as Bi$_2$Se$_3$\cite{zhang2009},  the ME response is isotropic, which can be described by the effective axion action $S_{\theta}$ with the quantized axion field $\theta=\pi$, protected by the time reversal symmetry $\mathcal{T}$. In AFM topological insulator MnBi$_2$Te$_4$, both the inversion symmetry $\mathcal{P}$ and $\mathcal{S}$ symmetry  constrain the axion field $\theta$ to the quantized value of $\pi$.  However, for even-SL MnBi$_2$Te$_4$ thin films, both $\mathcal{P}$ and $\mathcal{T}$ symmetries are broken and the mass term $m_5$ is induced, thus rendering unquantized $\theta$. It should be pointed out that due to the finite-size effect in MnBi$_2$Te$_4$ thin films, the ME response is generically anisotropic~\cite{liu2020tme}.  Moreover, since $m_5$ is related to the inherently fluctuating AFM order,  the axion field will be time dependent $\partial_t\theta(t)\neq0$, leading to the dynamical ME effect. In the following, we will show that the dynamical ME effect can be realized and drastically tuned by adjusting the thickness of MnBi$_2$Te$_4$ thin films, the temperature and element substitutions.

Here, we consider applying a magnetic field $\bm{B}$ in the plane of MnBi$_2$Te$_4$ thin films (e.g. the $x$ axis in Fig.~\ref{fig1}a), which will induce a charge polarization $\bm{P}$ in the same direction.
To describe such a ME response,  we need to calculate the ME coefficient $\alpha$ defined as $\bm{P}=-\alpha \bm{B}$ and it can be obtained from the quantity $\gamma\equiv(1/d)\int^{d/2}_{d/2}\eta(z)$ through $\alpha=\gamma(e^2/2h)$~\cite{liu2020tme, wang2015}. Here, $\eta(z)$ is a dimensionless function defined as 

\begin{equation}
\eta(z)=2\int_{-d/2}^{d/2}dz_1z_1\Pi_{xy}(z,z_1),
\end{equation}
where the current correlation function $\Pi_{xy}(z,z_1)$ is given by the Kubo formula, 
\begin{equation}
\begin{split}
\Pi_{xy}(z,z')=&\frac{\hbar^2}{2\pi e}\int d^2\bm{k}\sum_{n\neq m}f(\epsilon_{n\bm{k}})\\
\times&2\textrm{Im}\Big[\frac{\langle n\bm{k}|j_x^{3D}(\bm{k},z)|m\bm{k}\rangle\langle m\bm{k}|j_x^{3D}(\bm{k},z')|n\bm{k}\rangle}{(\epsilon_{n\bm{k}}-\epsilon_{m\bm{k}})^2}\Big].
\end{split}
\end{equation}
Here, $|n\bm{k}\rangle$ is the normalized Bloch wavefunction of the $n$-th electron subband, and $j^{3D}(\bm{k},z)=(e/\hbar)\partial_{\bm{k}}H_{3D}(\bm{k},z)$ represents the 3D in-plane current density operator, where $H_{3D}(\bm{k},z)$ is obtained by discretizing the system into a tight-binding model along the $z$ direction and fixing the periodic boundary condition in the $x$-$y$ plane to preserve $k_x$ and $k_y$ as good quantum numbers.

In the bulk limit of 3D topological insulators, the quantity $\gamma$ is related to the axion field by $\gamma=\theta/\pi$. According to Ref.~\cite{li2010dynamical},  the mass term $m_5$ could lead to a correction of $\theta$ to the linear order. Considering the inherent AFM fluctuations, which changes $m_5$ by $\delta m_5$, $\theta$ should be divided into a static part $\theta_0$ and a dynamical part $\delta\theta(t)$. As a result, $\gamma$ also consists of a static part $\gamma_0$ and a dynamical part $\delta\gamma(t)$ given by $\delta\gamma=\delta m_5/g$ with the coefficient $1/g=\partial\gamma/\partial m_5$ to the linear order of $\delta m_5$. We can see that $1/g$ is an important quantity for a large dynamical ME effect.

In Fig.~\ref{fig3}a, we present the quantity $\gamma$ as a function of the  thickness of $N$ bi-SLs MnBi$_2$Te$_4$ thin films (red line). It can be seen that with increasing $N$, $\gamma$ gradually increases and approaches 1 at the bulk limit ($\theta=\pi$) with $N\rightarrow\infty$. The growing deviation from $\gamma=1$ with decreasing $N$ results from the induced $m_5$ term by the increasing finite-size effect. We also calculated $\gamma$ of Mn$_2$Bi$_2$Te$_5$ films (yellow line) for the comparison. On the whole, the $\gamma$ has similar features for both MnBi$_2$Te$_4$ and Mn$_2$Bi$_2$Te$_5$ films due to the finite-size effect. The key difference is that $\gamma$ gradually approaches 1 for MnBi$_2$Te$_4$ films at the bulk limit but a finite value ($<1$) for for Mn$_2$Bi$_2$Te$_5$ films, which indicates the bulk MnBi$_2$Te$_4$ has $m_5=0$ ($\theta=\pi$) and the bulk Mn$_2$Bi$_2$Te$_5$ has $m_5\neq0$ ($\theta\sim\pi$). The dependence of $\gamma$ on the mass term $m_5$ for MnBi$_2$Te$_4$ films with different thickness ($N=4,6,8,16$), shown in Fig.~\ref{fig3}b. When $N$ is large, $\gamma$ quickly increases within the small $m_5$ region, and inversely when $N$ is small, $\gamma$ slowly increases. With further increasing $m_5$, $\gamma$ becomes to linearly decrease for all films, which indicates a large $m_5$ tends to suppress $\gamma$.

As for the dynamical part of $\gamma$, we calculated the quantity $1/g$ on different $m_5$ for $N$ bi-SL MnBi$_2$Te$_4$ films, shown in Fig.~\ref{fig3}d. It is found that $1/g$ increases with decreasing $m_5$, and the maximum value appears when $m_5\rightarrow 0$. In order to obtain a large quantity $1/g$, it needs a large $N$ with a small $m_5$. We know that the evolution of $m_5$ depends on the thickness $N$ of MnBi$_2$Te$_4$ films, shown in Figs.~\ref{fig2}c and \ref{fig2}d. Based on realistic values of $m_5$ induced by the finite-size effect, we then calculated the evolution of $1/g$ with varying $N$ in Fig.~\ref{fig3}c. We can see that $1/g$ has a large value for the ultra thin limit, and quickly  decreases with increasing the film thickness ($N$), and then crosses zero from positive to negative, and gradually saturates. The positive large  $1/g$ for the small $N$ indicates the increase of $\gamma$, and the saturation of $1/g$ for the large $N$ indicates the linear suppression, which is consistent with Fig.~\ref{fig3}b.  Importantly, stress that we can further tune $m_5\rightarrow0$ through raising the temperature to AFM N$\acute{e}e$l temperature to obtain a $drastically$ enhanced large $1/g$ of MnBi$_2$Te$_4$ films within a large tunable range for experiments, see the shaded region in Fig.~\ref{fig3}d. 

\begin{figure}[t]
  \centering
  \includegraphics[width=3.4in]{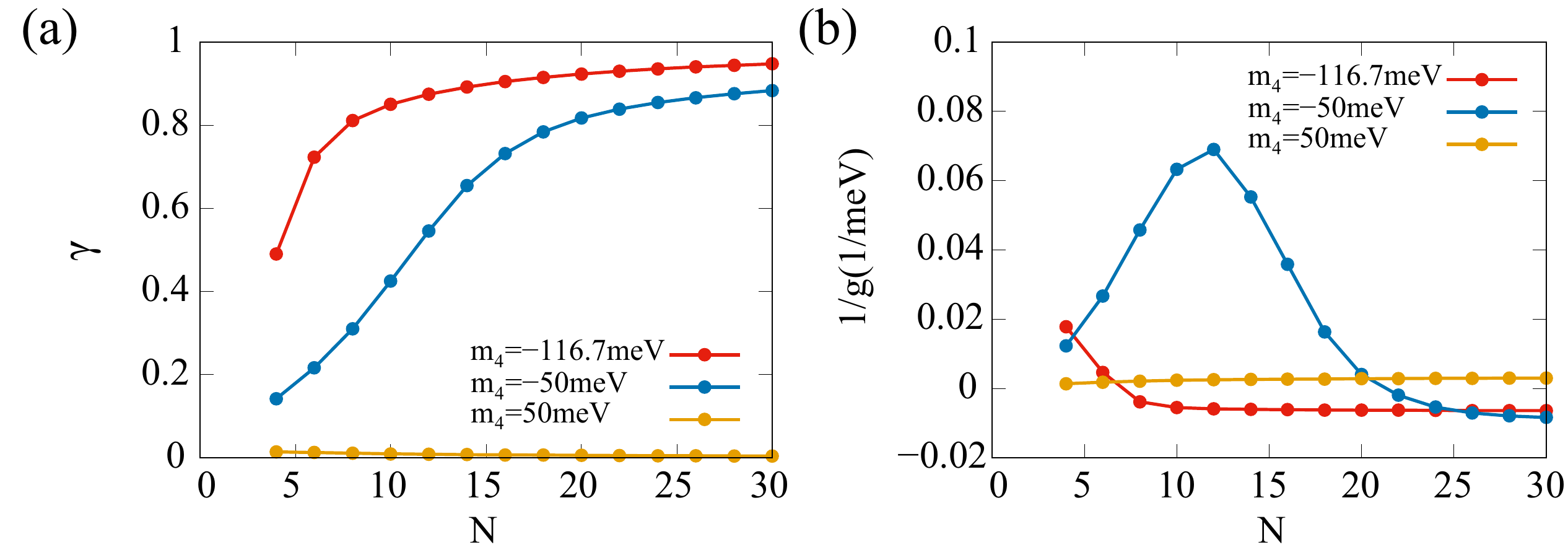}\\
  \caption{The magnetoelectric effects on $m_4$. The magnetoelectric quantity (a) $\gamma$ and (b) $1/g$ as a function of the thickness ($N$ bi-SLs) of films for three representative values of the mass term $m_{4}$, where negative (positive) $m_4$ represents topologically nontrivial (trivial) band structure with (without) the band inversion. The mass term $m_{4}$ can be tuned through the element substitution, for example, Mn(Bi$_{1-x}$Sb$_x$)$_2$(Te$_{1-y}$Se$_y$)$_4$.}\label{fig4}
\end{figure}

At last, we also investigate the dependence of the dynamical ME effect on the mass term $m_4$, with $m_4<0$ ($m_4>0$) indicating a topologically nontrivial (trivial) 3D electric structure. We choose three representative $m_4$ to show $\gamma$ and $1/g$ for different thickness $N$ of films in Fig.~\ref{fig4}. In the topologically trivial case ($m_4=50$ meV), $\gamma\sim0 $ ($\theta=0$). In the topologically nontrivial case ($m_4<0$), it can be seen that a larger $|m_4|$ exhibits a larger $\gamma$, which approaches $1$ ($\theta=\pi$) for $N\rightarrow\infty$. According to Fig.~\ref{fig4}b, $1/g$ in the topologically trivial case is quite small and obviously negligible compared to that in the topologically nontrivial case. Interestingly, $1/g$ can reach a maximum value through tuning $|m_4|$ and the thickness $N$, which also indicates that a proper band inversion ($|m_4|$) can give the largest $1/g$, not the strongest band inversion. Since $m_4$ is related to the band inversion, we can experimentally tune it through element substitutions in Mn(Bi$_{1-x}$Sb$_x$)$_2$(Te$_{1-y}$Se$_y$)$_4$ family materials. 

In conclusion, our work presents that AFM topological insulator MnBi$_2$Te$_4$ family films, which have been successfully synthesized and well studied, provide a perfect material platform to realize the dynamical ME effects. More importantly, the dynamical ME effect is tunable and even drastically enhanced through engineering the finite-size effect, the temperature and the element substitution. The exotic axion dynamics, including dynamical chiral magnetic effect, axionic polariton and nonlinear electromagnetic effect, should be experimentally realized in the proposed real-material platform. Our discovery in condensed matter physics might also be helpful to detect the real axion particle and the dark matter in the universe\cite{marsh2019}.

\emph{Methods.} We performed the first-principles calculations for bulk MnBi$_2$Te$_4$ and Mn$_2$Bi$_2$Te$_5$ of different spin-orbit coupling through employing the Vienna ab-initio simulation package(VASP) \cite{PhysRevB.54.11169, PhysRevB.47.4215} and the generalized gradient approximation (GGA) with the Perdew-Burke-Ernzerhof(PBE) \cite{PhysRevLett.77.3865, PhysRevB.50.17953} type exchange-correlation potential was adopted, with the energy cutoff fixed to 420 eV. The k-point sampling grid of Brillouin zone in the self-consistent process was a $\Gamma$-centered Monkhorst-Pack k-point mesh of $8\times8\times4$, and a total energy tolerance $10^{-7}$eV was adopted for self-consistent convergence. For MnBi$_2$Te$_4$ and Mn$_2$Bi$_2$Te$_5$ films, we employed the maximally localized Wanneir functions (MLWF) \cite{marzari1997maximally, souza2001maximally} to construct tight-binding Hamiltonians from the first-principle calculations. The Mn$-d$, Te$-p$ and Bi$-p$ orbitals were initialized for MLWFs by Wannier90 \cite{Pizzi2020}.

This work is supported by the Fundamental Research Funds for the Central Universities (Grant No. 020414380149) and the Natural Science Foundation of China (Grants No. 12074181, No. 11674165 and NO. 11834006) and the Fok Ying-Tong Education Foundation of China (Grant No. 161006).

\bibliography{daf}

\end{document}